\documentclass[useAMS,usenatbib,usegraphicx,rotating,color,multirow,upgreek]{aa}  
\usepackage{graphicx}
\usepackage{txfonts}
\usepackage{epsfig}
\usepackage{amsmath} 
\usepackage{rotating}           
\usepackage{color}     
\usepackage{graphicx}
\usepackage{upgreek} 

\def \HI {H{\sc \,i}}

\def\lapp{\ifmmode\stackrel{<}{_{\sim}}\else$\stackrel{<}{_{\sim}}$\fi}
\def\gapp{\ifmmode\stackrel{>}{_{\sim}}\else$\stackrel{>}{_{\sim}}$\fi}

\begin{document} 
\title{QSO redshift estimates from optical, near-infrared and ultraviolet colours}
 \author{S. J. Curran
           \and
           J. P. Moss
         }
  \institute{School of Chemical and Physical Sciences, Victoria University of Wellington, PO Box 600, Wellington 6140, New Zealand\\
  \email{Stephen.Curran@vuw.ac.nz} 
}

\abstract{ A simple estimate of the photometric redshift would prove invaluable to forthcoming continuum surveys on the
  next generation of large radio telescopes, as well as mitigating the existing bias towards the most optically bright
  sources.  While there is a well known correlation between the near-infrared $K$-band magnitude and redshift for
  galaxies, we find the $K-z$ relation to break down for samples dominated by quasi-stellar objects (QSOs). We
  hypothesise that this is due to the additional contribution to the near-infrared flux by the active galactic nucleus
  (AGN), and, as such, the $K$-band magnitude can only provide a lower limit to the redshift in the case of active
  galactic nuclei, which will dominate the radio surveys. From a large optical dataset, we find a tight relationship
  between the rest-frame $(U-K)/(W2-FUV)$ colour ratio and spectroscopic redshift over a sample of 17\,000 sources,
  spanning $z\approx0.1 - 5$.  Using the {\em observed}-frame ratios of $(U-K)/(W2-FUV)$ for redshifts of $z\lapp1$,
  $(I-W2)/(W3-U)$ for $1\lapp z \lapp3$ and $(I-W2.5)/(W4-R)$ for $z \gapp3$, where $W2.5$ is the $\lambda=8.0$~$\mu$m
  magnitude and the appropriate redshift ranges are estimated from the $W2$ (4.5~$\mu$m) magnitude, we find this to be a
  robust photometric redshift estimator for quasars. We suggest that the rest-frame $U-K$ colour traces the excess flux
  from the AGN over this wide range of redshifts, although the $W2-FUV$ colour is required to break the degeneracy.}
  
   \keywords{techniques: photometric  -- methods: statistical --  galaxies: active --  galaxies: photometry -- infrared: galaxies -- ultraviolet: galaxies}

   \maketitle
%

\section{Introduction}

Continuum surveys with forthcoming large radio telescopes, the {\em Square Kilometre Array} (SKA) and its pathfinders,
are expected to yield vast numbers of new sources for which an estimate of the redshift will prove invaluable in
extending the scope of the science outcomes. For example, the {\em Evolutionary Map of the Universe} (EMU,
\citealt{nha+11}) on the {\em Australian Square Kilometre Array Pathfinder} (ASKAP), 
will take a census of 70 million radio sources in the sky. Being a continuum survey, the spectra will be of
insufficient resolution to determine the source redshifts. If, however, these can be reliably estimated from the
photometry alone, the value of the survey in determining how the Universe is populated will increase dramatically.

Even where wide-band radio spectroscopy is available, via 21-cm absorption of neutral hydrogen (\HI), an independent
measure of the redshift will allow us to determine whether the absorbing gas is located within the host of the continuum
source or arises in some intervening system. For example, recent detections of \HI\ 21-cm absorption with the six
antenna {\em Boolardy Engineering Test Array} of ASKAP have required follow up observations on large optical
instruments in order to ascertain whether they are associated with  or intervene the background source
\citep{asm+15,asm+15a,amm+16}. This is important in determining the populations of active and quiescent sources in the distant Universe and will provide a valuable complement to machine learning methods \citep{cdda16}.  
On the full 36 antenna ASKAP, the {\em First Large Absorption Survey in \HI} (FLASH) is expected to
yield spectra for 150\,000 radio sources, and so, observationally expensive, optical spectroscopy is not practical, an
issue which will be more severe for the SKA \citep{msc+15}.

Having an estimate of the redshift, to which to tune the receiver, without the reliance upon an optical spectrum, is also
desirable for current high redshift decimetre and millimetre band spectral line surveys.  Specifically, sources which
are sufficiently bright to yield a reliable optical redshift bias against the most dust-rich systems: In both
intervening and associated systems the strength of the absorption is correlated with the red colour of the source
\citep{wfp+95,cmr+98,cwm+06}, suggesting that the reddening is due to dust, which shields the neutral gas from the
ambient UV field. Furthermore, at high redshift, visual magnitudes of $\lapp23$ correspond to $Q_\text{\HI}\gapp 10^{56}$
ionising photons per second in the source frame\footnote{For instance, $R\lapp21$ \citep{cwsb12} and $B\lapp23$ \citep{cwt+12} give $Q_\text{\HI}\gapp 10^{56}$~s$^{-1}$ at $z\gapp3$.}, which is sufficient to ionise all of the neutral
gas within the host galaxy \citep{cww+08,cw12}. 
This suggests that even the SKA will not detect the star forming reservoir 
in the currently known high redshift radio sources for which we have an optical redshift. Therefore, some other means, by
which the redshift can be estimated for  fainter objects, is required.

\begin{figure}
\centering \includegraphics[angle=-90,scale=0.5]{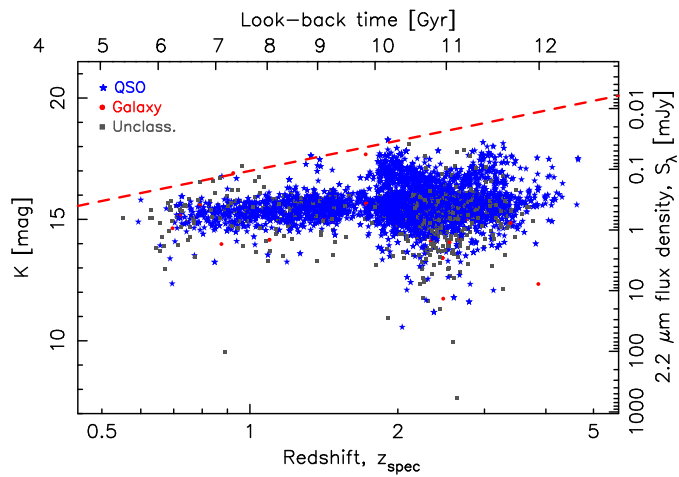}
\caption{The Hubble $K$-band diagram for the initial sample (see Sect.~\ref{sec:samp}). The stars signify QSOs, the circles galaxies and squares unclassified sources. The broken line shows the fit of \citet{dvs+02}.}
\label{K-z_MgII}
\end{figure}

There are numerous methods used to estimate photometric redshifts (see \citealt{sih18}), although these
can be very complex (\citealt{nsl+19}, see also Sect. \ref{cwos}).
Perhaps the simplest is the strong
relationship between the near-infrared $K$-magnitude ($\lambda = 2.2~\mu$m) and the redshift of the source
\citep{dvs+02,wrjb03}.
Even though more distant objects will be fainter, the narrow
spread of this correlation is nevertheless remarkable, given that each source will have its own intrinsic luminosity. This only
applies to galaxies, however, and when QSOs are added the relationship is lost
(Fig.~\ref{K-z_MgII}).
The fact that the sources move to the right of the $K = 4.43\log_{10}z +17$ fit \citep{dvs+02}, demonstrates that the $K$-magnitude underestimates the redshift in the case of QSOs. In other words, at a given redshift, a QSO is brighter in near-infrared (NIR) emission than a galaxy, most likely due to the contribution of the active galactic nucleus. 
Thus, the best the $K$-magnitude can do is provide a lower limit to the redshift. In this paper, we present our
efforts to account for the AGN contribution, thus providing a reliable photometric redshift estimator of QSOs, over
a range of selection criteria and  independent of many of the assumptions required by other methods.

\section{Analysis and results}

\subsection{Photometry matching}
\label{pm}

\begin{figure}
\centering \includegraphics[angle=-90,scale=0.52]{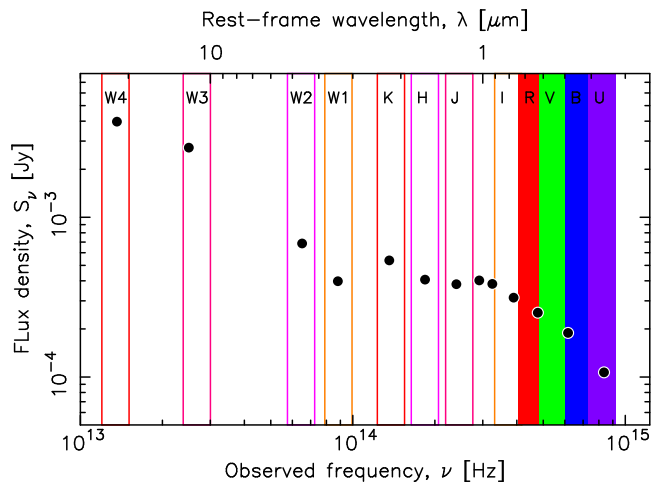}
\caption{Example of the photometric data points from a single source. The canonical band
frequencies with $\Delta\log_{10}\nu=\pm0.05$ are shown.}
\label{poly_fit}
\end{figure}

For each source, we matched the coordinates to the closest source within a 6 arc-second search radius in the {\em
  NASA/IPAC Extragalactic Database} (NED), from which we obtained the specific flux densities. We also used the NED names
to query the {\em Wide-Field Infrared Survey Explorer} (WISE, \citealt{wem+10}), the {\em Two Micron All Sky Survey}
(2MASS, \citealt{scs+06}) and the {\em Galaxy Evolution Explorer} (GALEX data release
GR6/7)\footnote{http://galex.stsci.edu/GR6/\#mission} databases.  In order to ensure a uniform magnitude measure, if the
frequency of the photometric point fell within $\Delta\log_{10}\nu=\pm0.05$ of the central frequency of the
band the measurement was added (Fig.~\ref{poly_fit}).\footnote{This method was also used in Fig.\ref{K-z_MgII}.}
For more than one point in the band the fluxes were averaged before the conversion to magnitude.

\begin{figure}
\centering \includegraphics[angle=-90,scale=0.5]{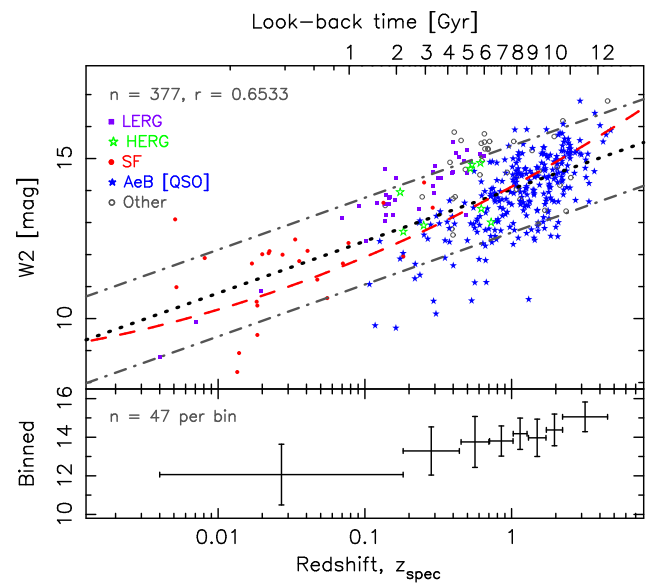}
\caption{The Hubble $W2$-band diagram for the LARGESS sample. The dotted line shows the least-squares fit, with
  regression coefficient $r$, and the dot--dashed lines showing the $\pm1\sigma$ span of this.  The filled squares signify
  high-excitation radio galaxies (HERGs), the unfilled stars low-excitation radio galaxies (LERGs), the circles
  star-forming (SF) galaxies and the filled stars broad emission line (AeB) sources (suspected quasars,
  \citealt{csc+17}). The broken curve shows the fit of \citet{gas+18} to the QSOs (their Table~2).  The bottom panel
  shows the mean values in equally sized bins with the vertical error bars showing $\pm1\sigma$ and the horizontal bars
  the range.}
\label{W2_LARGESS}  
\end{figure}

\begin{figure}
\centering \includegraphics[angle=-90,scale=0.5]{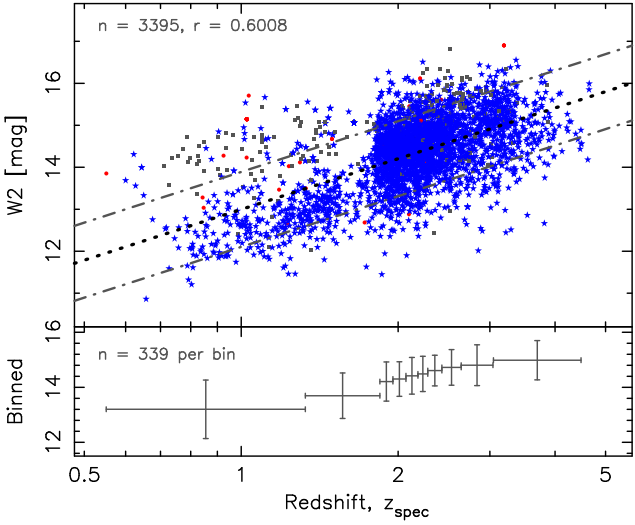}
\caption{As Fig.~\ref{W2_LARGESS}, but for our initial (the MgII) sample (see main text). Using $W1$ gives $r = 0.4895 (n=3400)$. 
As per Fig.~\ref{K-z_MgII}, the stars signify QSOs, the circles galaxies and squares unclassified sources.}
\label{W2_MgII}
\end{figure}

Using the {\em Large Area Radio Galaxy Evolution Spectroscopic Survey} (LARGESS, \citealt{csc+17}), \citet{gas+18} find
a correlation between redshift and the $W1$ ($\lambda=3.4$~$\mu$m) and $W2$ ($\lambda=4.6$~$\mu$m) magnitudes of WISE, which includes quasars (or at least, broad emission
line sources). However, the spread is wide, with \citeauthor{gas+18} quoting regression coefficients of $r = 0.56$ and
0.36 for the $W1$ and $W2$-band fits, respectively. From our own matching, by source (NED) name, we obtain $r = 0.77$ and
0.65, respectively (Fig.~\ref{W2_LARGESS}).\footnote{This yielded far fewer (377, Fig.~\ref{W2_LARGESS}) sources
with a $W2$ measure than the method of \citet{gas+18}, which matches 9\,294 of the LARGESS
sources.}
Although indicative of reasonable fits, when applied to our initial  test sample, the ``MgII sample'' (Fig.~\ref{W2_MgII}),
the regression coefficients drop, indicating that the $W1$ and $W2$ fits may not prove to be reliable photometric
redshift predictors for other samples.

\subsection{The  MgII  sample and initial testing}
\label{sec:samp}

Our ultimate aim is to test the 3.3 million galaxies and QSOs in the {\em Sloan Digital Sky Survey} (SDSS) Data Release
12 (DR12, \citealt{aaa+15}), which we are currently querying for the full NED, WISE, 2MASS and GALEX photometries.
This is expected to take several years to complete and so we initially
tested the 23\,659 sources illuminating MgII absorbers \citep{zm13} in the SDSS DR12.  Of these, photometric data
could be found for 17\,285.

In order to find the best combination of magnitudes with which to obtain an estimator
for the photometric redshift, we initially explored machine learning techniques. In the
{\sc weka} package \citep{hfh+09}, a suite of machine learning algorithms,
only individual features (magnitudes) could be tested
automatically with combinations having to be entered manually as features.\footnote{See \citet{cdda16,cd18} 
for examples.} We therefore proceeded manually, performing exhaustive tests of various arithmetic
combinations of magnitudes and colours. Although many of these may not have
been meaningful (e.g. the multiplication of magnitudes), they were, nevertheless, tested as they
may provide insight on how to proceed \citep{mos19}.

\begin{figure}
\centering \includegraphics[angle=-90,scale=0.5]{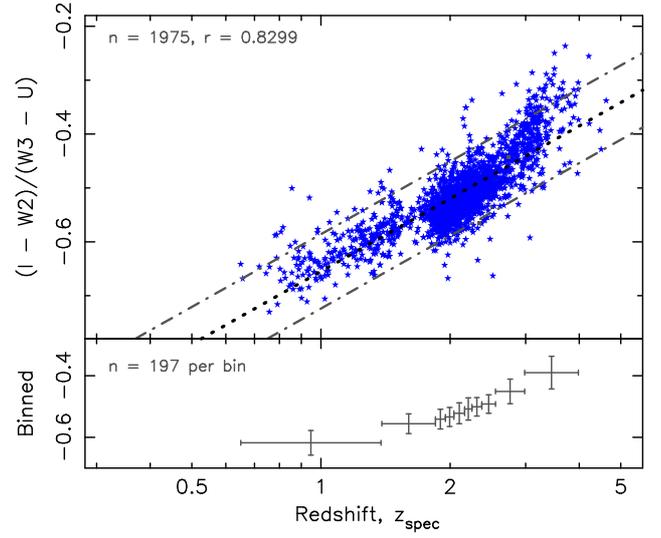}
\caption{The best fit model to the MgII sample. The dotted line shows the least-squares fit, giving $(I-W2)/(W3-U)
  = 0.448\log_{10}z -0.655$.}
\label{I-W2overW3-U}
\end{figure}

Since small samples could yield large regression coefficients, in order to be flagged as a good fit, both $|r| > 0.5$ and
$r^2n>1000$ were required.  For the magnitude combinations, the best fit was given by $U-I$ with $r = 0.5887$ and $r^2n
= 2340$. 
Of the colour combinations, there were several with $|r| > 0.7$ and $r^2n>1000$, with $(I-W2)/(W3-U)$ being one of the
top five with $r = 0.83$ and $r^2n = 1360$ (Fig.~\ref{I-W2overW3-U}).
The remaining four, which had slightly lower values of $|r|$, but slightly higher sample sizes, giving a higher $r^2n$,
had a complex physical interpretation, e.g. $(B\times W2)/(R \times I)$, whereas $(I-W2)/(W3-U)$ is 
recognisable as a colour--colour relation.

\subsection{SDSS DR12 -- first 50\,000 QSOs}
\label{dr12}

\label{soc}

\begin{figure}
\centering \includegraphics[angle=-90,scale=0.5]{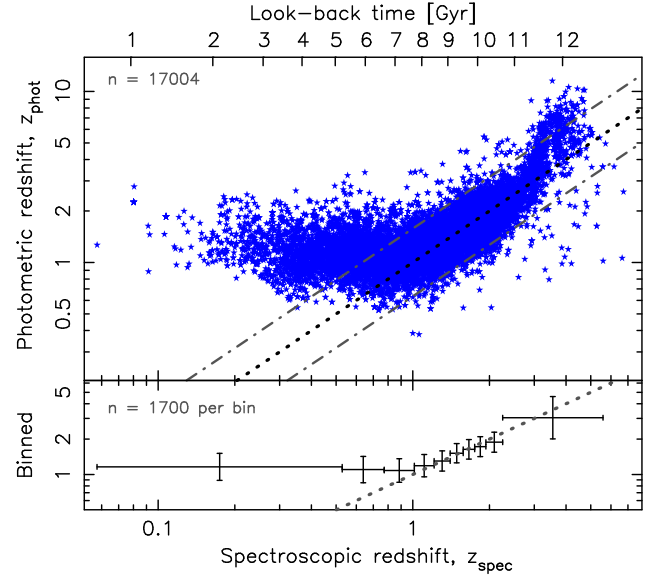}
\caption{The photometric redshifts predicted for the SDSS sample. The dotted line shows $z_{\rm pred} = z_{\rm spec}$, with the dot--dashed lines showing the $\pm1\sigma$ range.}
\label{SDSS_plain}
\end{figure}

While testing the initial sample, the data-mining of the SDSS DR12 was ongoing 
and we now discuss the first 50\,000 QSOs with accurate spectroscopic redshifts ($\delta z/z<0.01$).
In  Fig.~\ref{SDSS_plain}, we show the photometric redshifts predicted for these from the MgII model.
Although the fit is similar over the range covered by the MgII sample, this fails at $z\lapp1$. 

\begin{figure*}
\centering \includegraphics[angle=-90,scale=0.62]{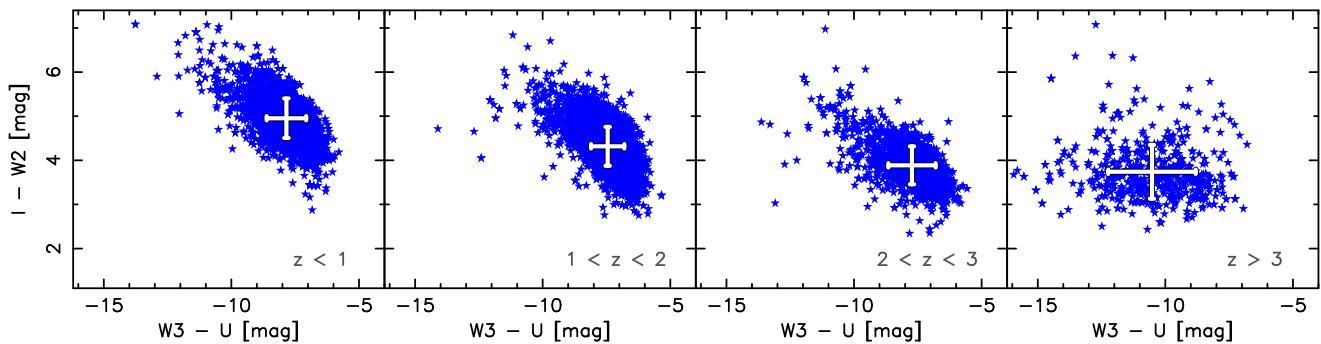}
\caption{The $(I-W2)/(W3-U)$ colour--colour plot over various redshift ranges. The error bars show the $\pm1\sigma$ ranges around the mean.}
\label{versus}
\end{figure*}

\begin{figure}
\centering \includegraphics[angle=-90,scale=0.52]{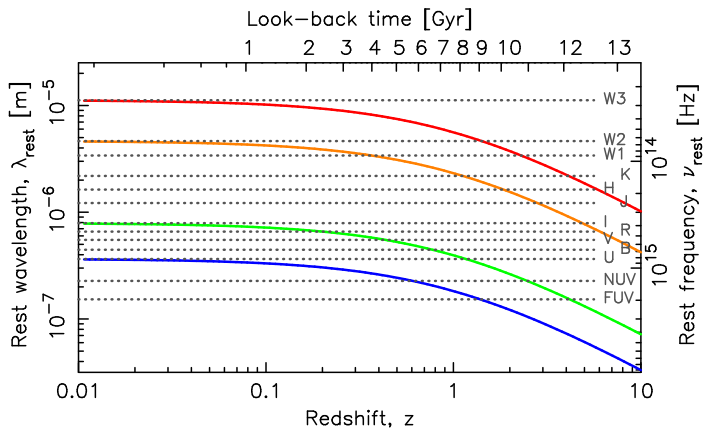}
\caption{The evolution of source-frame wavelength/frequency with redshift for $W3, W2, I$ and $U$. These are overlaid
  upon the observed-frame near-infrared, optical and GALEX near-ultraviolet (NUV) and far-ultraviolet (FUV) bands.}
\label{evolv}
\end{figure}

The $W3, W2, I$ and $U$ bands have central wavelengths of 12, 4.6, 0.806 and 0.365~$\mu$m, respectively, 
and so $I-W2$ and $W3-U$ trace the extreme red --- near-infrared  and near-infrared --- ultraviolet colours,
respectively. Looking at the colours individually (Fig.~\ref{versus}), 
both $I-W2$ and $W3-U$ decrease with redshift, although it is not clear why these should be particularly sensitive to
the source redshift. However, it should be borne in mind that at $z\gg0$ these would have been emitted at significantly
shorter wavelengths.  For instance, at a redshift of $z\sim1$, the observed $I-W2$ and $W3-U$ colours are $U-K$ and $W2
- FUV$ in the rest-frame of the source, respectively (Fig.~\ref{evolv}).

\begin{figure*}
\centering \includegraphics[angle=-90,scale=0.4]{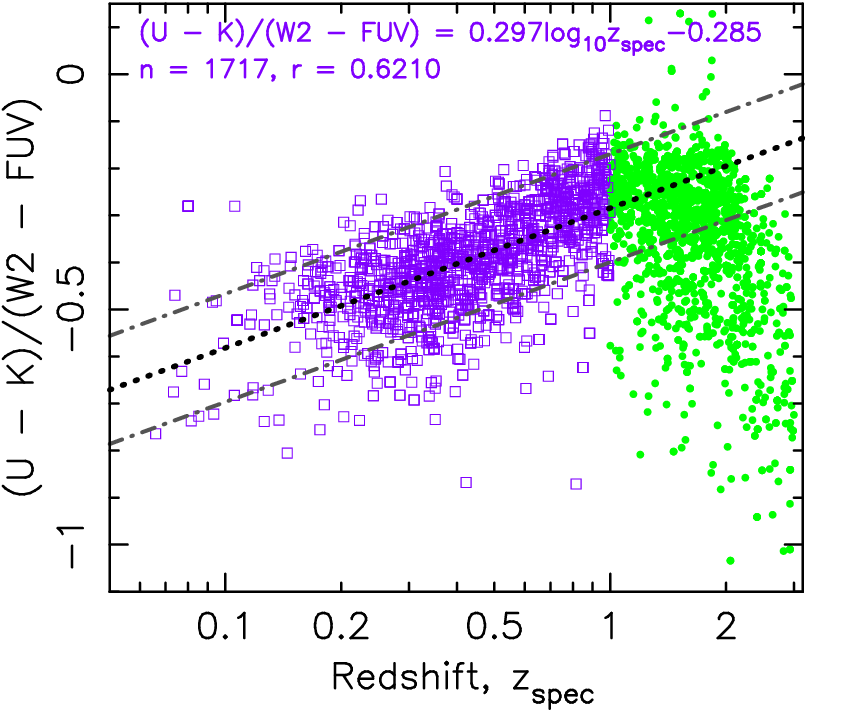}
\centering \includegraphics[angle=-90,scale=0.4]{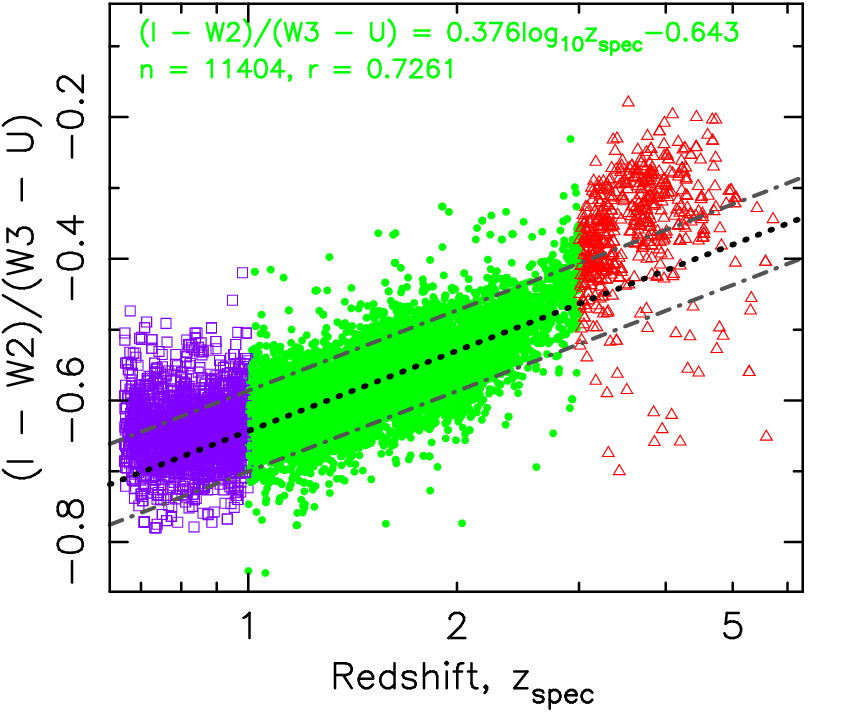}
\centering \includegraphics[angle=-90,scale=0.4]{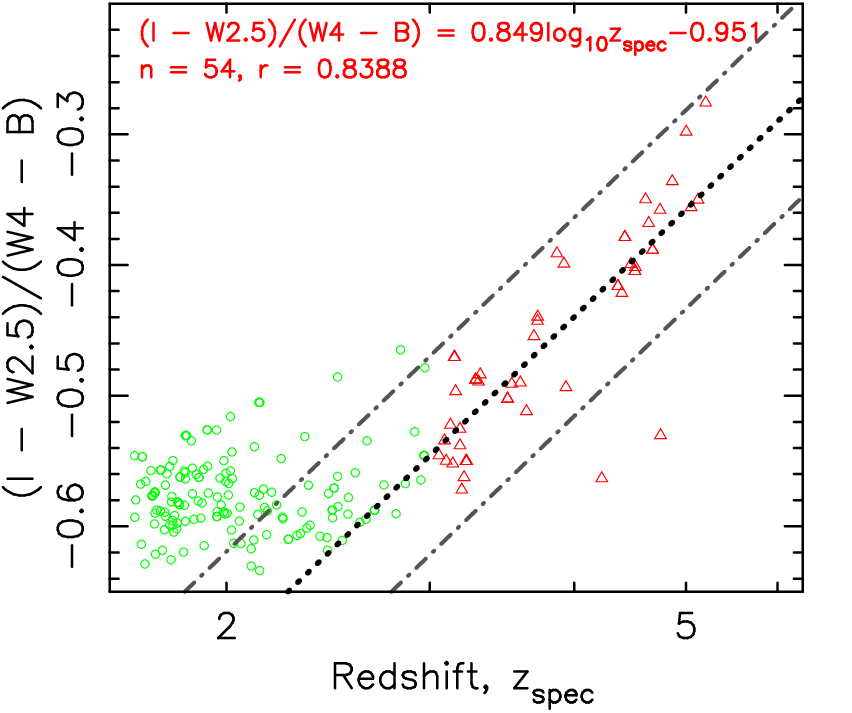}
\caption{As Fig.~\ref{SDSS_plain},  fitting the observed-frame $(U-K)/(W2-FUV)$ for $z<1$ (squares, left), $(I-W2)/(W3-U)$ for $1\leq z \leq3$ (circles, middle) and $(I-W2.5)/(W4-R)$ for $z>3$ (triangles, right).}
\label{side}
\end{figure*}

In order to fit the $z\lapp1$ data, we tested further combinations of magnitudes, although, as expected
from above, the combination $(U-K)/(W2 - FUV)$ gives the best result (Fig.~\ref{side}, left).
In Fig.~\ref{SDSS_plain}, we also note a departure from the $z_{\rm spec}\gapp1$ trend at $z_{\rm spec}\gapp3$, also
evident as the increased scatter in the $z>3$ panel of Fig.~\ref{versus}.  At these redshifts, the rest-frame $(U-K)/(W2
- FUV)$ combination will be approximately $(J-W2.5)/(W4 - V)$ in the observed-frame, where we have dubbed the $\lambda=8.0$~$\mu$m magnitude,
located between the $W2$ (4.5~$\mu$m) and $W3$ (12~$\mu$m) bands, ``$W2.5$''.\footnote{The {\em Spitzer Space Telescope}
  \citep{ctb+13} data is incorporated into the NED photometry search (Sect.~\ref{sec:samp}).}  Possibly due to the
limited data, it is in fact the observed $(I-W2.5)/(W4 - B)$ which gives the tightest correlation with $z_{\rm spec}$, although the
numbers remain small (Fig.~\ref{side}, right).

An issue with using a multi-component fit is deciding where the combination should be switched in the absence of a
spectroscopic redshift.  One possibility is to use the photometric redshift (e.g. \citealt{mhp+12}), although, due to
the flattening of the $z_{\rm phot}$---$z_{\rm spec}$ relation at $z_{\rm phot}\sim1$, this is
not helpful with $z_{\rm spec}\approx0.1 -2$ being spanned for $z_{\rm phot}\approx1$  (Fig.~\ref{SDSS_plain}). Various methods, which did not rely upon
knowledge of the spectroscopic redshift, were trialled, but were unsuccessful. For example, using the value of
$(U-K)/(W2 - FUV)$, but, as seen from (Fig.~\ref{side}, left), the variation of this combination with $z_{\rm spec}$
exhibits a turnover at $z_{\rm spec}\approx1$, resulting in a degeneracy. The most straightforward and effective method
was a switch at $W2\approx12.5$ for $(U-K)/(W2 - FUV)$ to $(I-W2)/(W3-U)$ and at $W2\approx15$ for
$(I-W2)/(W3-U)$ to $(I-W2.5)/(W4 - B)$, based upon the loose $W2$---$z_{\rm spec}$ relation (Fig.~\ref{W2_MgII}). 

\begin{figure}
\centering \includegraphics[angle=-90,scale=0.52]{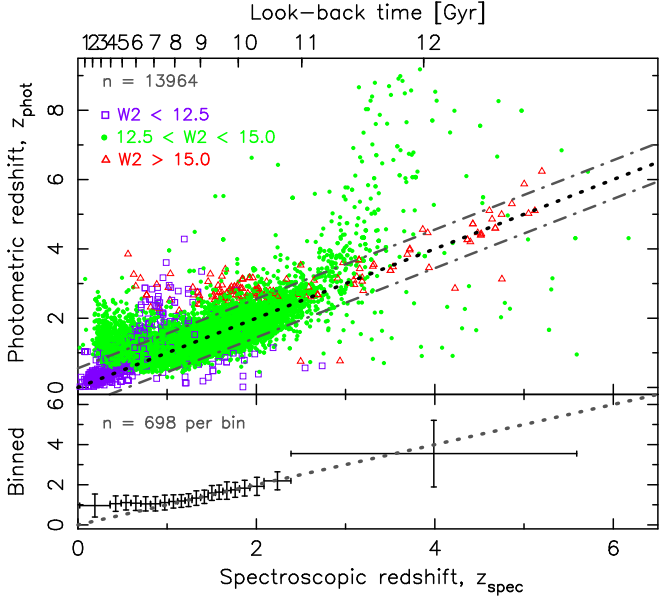}
\caption{The photometric redshifts predicted from the multi-component fit to the SDSS data (Fig.~\ref{side}), 
shown on a linear scale to emphasise the high redshift range. The symbols are as per Fig.~\ref{side}, in order
to show the $W2$ switch of each point.}
\label{lin}
\end{figure}

In Fig.~\ref{lin}, we show the resulting photometric redshifts obtained from the fits in Fig.~\ref{side}.
As seen from this, there is some scatter at  $z_{\rm spec} \lapp0.5$, due to the
imperfect switch invoking the $W2$ magnitude, and at $z_{\rm spec} \gapp3$, again, possibly due to an imperfect
switch, in addition to the sparsely sampled data (Fig.~\ref{side}, right).
However, from $z\gapp0.5$ up to the redshift limit of the data, the model is seen to give accurate statistical
predictions of the photometric redshift.

\section{Discussion}

\subsection{Application to radio selected samples}
\label{rss}

Our model is constructed from an optically selected sample and here we test it upon several
radio band datasets, which cover a wide range of redshifts. By using  disparate and 
heterogeneous test samples, we hope to maximise the robustness of our model to other (future) datasets
where information may be limited. 

The LARGESS catalogue comprises 19\,179 radio sources matched with SDSS counterparts, giving redshifts for 10\,883  \citep{csc+17}.
Upon removing duplicate sight-lines, the $U\cap K \cap W2 \cap FUV$, $W3\cap W2 \cap\, I\cap U$ and $I\cap W2.5 \cap\,
W4\cap B$ requirements to cover all redshifts yielded only 250 sources, which is 2.3\% of the sources with known
redshifts. This compares to 85.6\% for the $W2$ magnitude only (\citealt{gas+18}), although our more stringent matching of sources
will also contribute to the small numbers (see Sect.~\ref{sec:samp}).
\begin{figure}
\centering \includegraphics[angle=-90,scale=0.5]{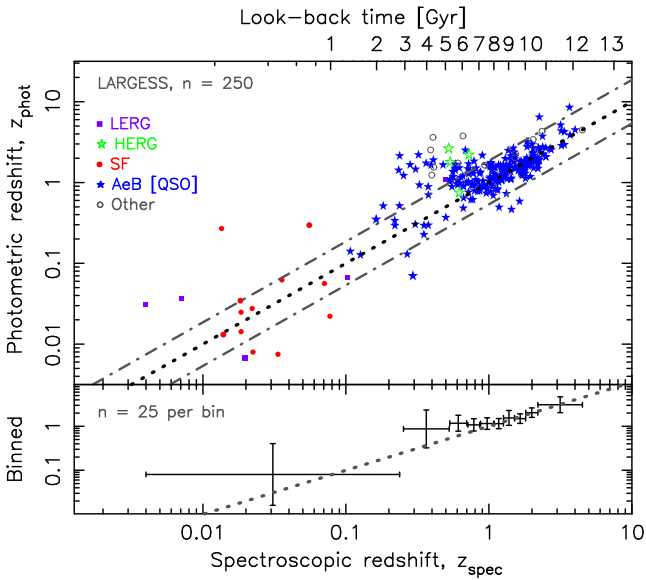}
\caption{The predicted redshifts for the LARGESS sample.}
\label{LARGESS_pred}
\end{figure} 
From the fit (Fig.~\ref{LARGESS_pred}), we see that our model provides reasonable photometric redshifts for both AGN and
non-AGN, deviating by a maximum of $\approx1\sigma$, at $z_{\rm spec}\approx0.5$, close to where the observed $(U-K)/(W2 - FUV)$ to $(I-W2)/(W3-U)$ switch occurs.

The {\em Second Realization of the International Celestial Reference Frame by Very Long Baseline Interferometry} (ICRF2,
\citealt{mab+09}), constitutes a sample of strong flat spectrum radio sources, of which 1682 have known redshifts
(\citealt{tm09,tsj+13} and references therein). Of these 119 (8.0\% of the sample)\footnote{Out of 1486, upon the removal of duplicates and unreliable redshifts.} have all of the required magnitudes.
\begin{figure}
\centering \includegraphics[angle=-90,scale=0.5]{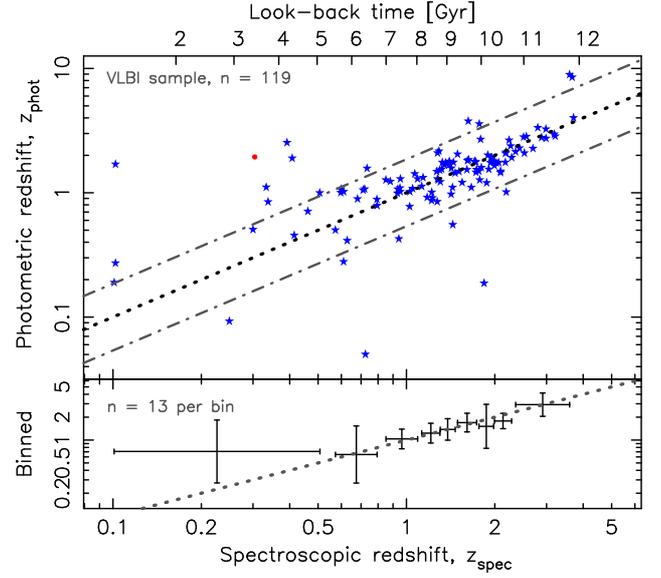}
\caption{As Fig. \ref{LARGESS_pred}, but for the VLBI sample.  As per Fig.~\ref{K-z_MgII}, the stars signify QSOs, the circles galaxies and squares unclassified sources.}
\label{VLBI_pred}
\end{figure} 
Although the dataset is small (Fig.~\ref{VLBI_pred}), accurate photometric redshifts are predicted for
$z_{\rm spec}\gapp0.5$, below which the data are sparser.

In addition to the ICRF2, there are a multitude of radio source catalogues. However, these are generally lacking in
spectroscopic information.  Of those for which redshifts exist, the {\em Combined ES-NVSS Survey Of Radio Sources}
(CENSORS) tested by \citet{gas+18} has 143 redshifts \citep{bbp+08} and the {\em GaLactic and Extragalactic All-sky
  Murchison Widefield Array} (GLEAM) survey has 215 redshifts \citep{ceg+17}.\footnote{Although very few could be
  matched to within 6 arc-seconds of a NED source (see Sect.~\ref{pm}).}  The Parkes Flat-Spectrum samples also have
measured redshifts; the {\em Parkes Half-Jansky Flat-Spectrum Sample} (PHFS) with 277 (\citealt{dwf+97}) and the {\em
  Parkes Quarter-Jansky Flat-spectrum Sample} (PQFS) with 470 (\citealt{jws+02}).\footnote{From 49 GHz peaked spectrum
  radio galaxies in the PHFS, \citet{vss+07} find a correlation between the $R$-band magnitude and the
  redshift. However, none of these sources has the full $W3, W2, I, U$ combination.}
Since these samples are relatively small, we use the redshifted radio sources searched in associated \HI\
21-cm absorption. This comprises 819 sources over redshifts of $0.002 \leq z \leq 5.19$ (see \citealt{cd18,chj+19}
and references therein), in addition to providing a more heterogeneous (unbiased) sample than the aforementioned catalogues.
\begin{figure}
\centering \includegraphics[angle=-90,scale=0.5]{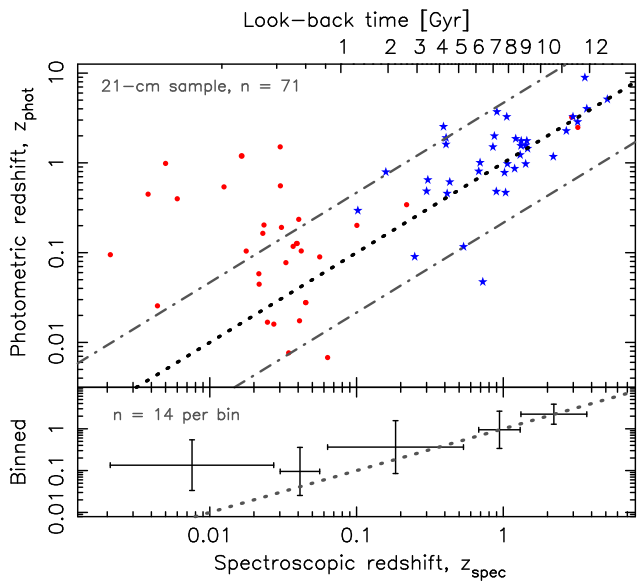}
\caption{As Fig. \ref{VLBI_pred}, but for the radio sources searched in \HI\ 21-cm absorption.}
\label{21-cm_pred}
\end{figure} 
Of these, the required magnitudes could be measured for 71 sources, giving a fraction of 8.7\%. Again, the
predicted photometric redshifts are statistically accurate down to redshifts of $z\lapp0.03$,
where galaxies dominate (Fig.~\ref{21-cm_pred}).

\subsection{Comparison with other studies}
\label{cwos}

\begin{figure}
\centering \includegraphics[angle=-90,scale=0.52]{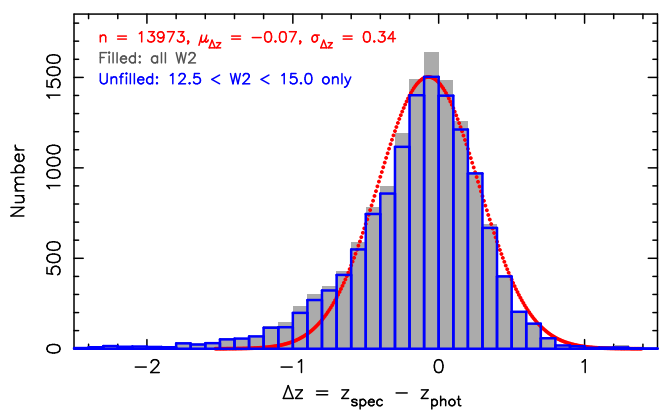}
\caption{The distribution of $z_{\rm spec} - z_{\rm phot}$ over all redshifts with no source filtering. The unfilled
  histogram shows the $12.5 < W2 < 15.0$ [observed-frame $(I-W2)/(W3-U)$] data only and the filled all of the data [i.e. including
  the observed-frame $(U-K)/(W2 - FUV)$ and $(I-W2.5)/(W4-R)$ data], to which the Gaussian is fit.}
\label{histo}
\end{figure}

In Fig.~\ref{histo}, we show the distribution of
$\Delta  z = z_{\rm spec} - z_{\rm phot}$, which  appears to be well fit by a Gaussian,
apart from the  extended tail at $\Delta z\lapp-0.7$.
We note also, that the different fits over the three $W2$ ranges (Fig.~\ref{side}) give consistent results.

\begin{figure}
\centering \includegraphics[angle=-90,scale=0.52]{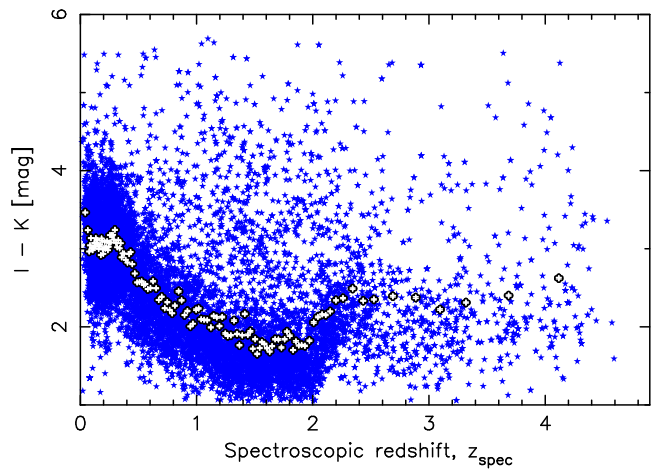}
\caption{The $I-K$ colour--redshift relation for our sample (cf. figure 13 of \citealt{mhp+12}). The overlain error bars show the mean values in equally sized bins (of 100), spanning the
error range ($\pm1\sigma/10$).} 
\label{I-K}
\end{figure}

Other studies have also used (earlier releases of) SDSS data to yield narrower distributions of $\Delta z$, typically being 
$\sigma_{\Delta z}\approx0.1$, with 70\% of the values being within $|\Delta z|\approx 0.2$ \citep{rws+01,wrs+04,bbm+08,mhp+12}, whereas we require $|\Delta z|= 0.4$ to reach this fraction.
These Gaussian fits are, however, considerably narrower than the distributions, which exhibit wide tails on both sides
(see Appendix \ref{appa}).
Furthermore, the methods employed by these studies are considerably more complex than the method presented here,
invoking the evolution of the SDSS colours (e.g. \citealt{rws+01}) or $i -K$ \citep{mhp+12} with redshift (Fig.~\ref{I-K}).
This introduces a degeneracy, where a colour matches more than one spectroscopic redshift, requiring the
use of specialised algorithms to break this. The test samples are also filtered,  with the removal of redder
sources (e.g. \citealt{rws+01}) and the visual inspection of images being required before their inclusion (e.g. \citealt{mhp+12}).
Lastly, the tight $z_{\rm spec}$---$z_{\rm phot}$ relationships are obtained over  limited (photometric) redshifts $0.8 \leq z_{\rm phot} \leq 2.2$ 
\citep{wrs+04}, $1.0 \leq z_{\rm phot} < 3.5$ \citep{mhp+12} and magnitudes \citep{bbm+08}. 

\begin{figure}
\centering \includegraphics[angle=-90,scale=0.52]{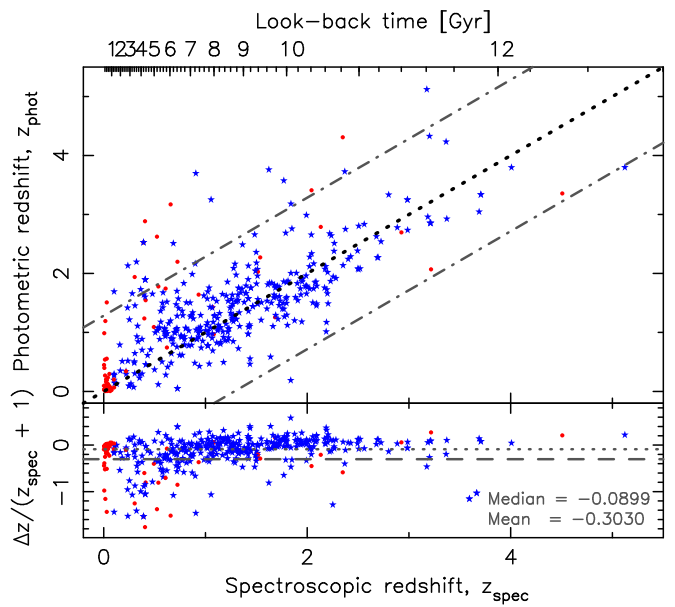}
\caption{All of the radio selected sources (Figs.\ref{LARGESS_pred} to \ref{21-cm_pred}) shown on a linear scale. The bottom panel
shows the normalised residuals, with the dotted line showing the median value and the dashed line the mean.}
\label{radio_lin}
\end{figure}

Predicting the photometric redshifts of radio selected data has been explored by \citet{lnp18}, who,
through machine learning techniques, find $\approx90$\% of the photometric redshifts to lie within 
a normalised residual of $\Delta z/(z_{\rm spec}+1) \leq\pm0.15$. From our radio sample, which with 441 sources is of a similar size
to the \citeauthor{lnp18} test samples (281--855), we find that only 60\% of the sources have $\Delta z/(z_{\rm spec}+1) \leq\pm0.15$ (Fig.~\ref{radio_lin}), with 90\% being reached at $\Delta z/(z_{\rm spec}+1) \lapp\pm0.5$.
However, the \citeauthor{lnp18} data are concentrated at $z_{\rm spec}\lapp0.5$, where the scatter being mitigated by the large $z_{\rm spec}$ values in the denominator of the normalised residual, whereas 
 the few data points at $z_{\rm spec}\gapp1$ have $\Delta z/(z_{\rm spec}+1) >> \pm0.15$.

\begin{figure}
\centering \includegraphics[angle=-90,scale=0.52]{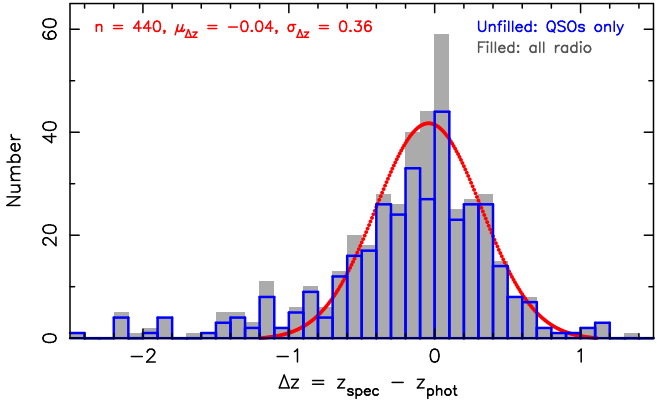}
\caption{The distribution of $z_{\rm spec} - z_{\rm phot}$ over all redshifts with no source filtering for the 
radio selected samples. The unfilled histogram shows the QSOs (quasars) only and the filled all of the data, to which the Gaussian is fit.}
\label{radio}
\end{figure}

Lastly, we summarise $\Delta z$ for our radio selected sources in Fig.~\ref{radio}.
Although the numbers are small, the distribution is very similar that obtained from the optical data (Fig.~\ref{histo}),
with an almost identical width  and $\Delta z \lapp-0.7$ tail. We note also that, despite the apparent scatter introduced
by the galaxies in the 21-cm sample (Fig.~\ref{21-cm_pred}), these follow a similar distribution as the QSOs.

\subsection{Magnitude limitations}
\label{maglim}

Although the results are very promising, with the photometric redshifts of the radio selected sources being as accurate
as from the optically selected parent sample, the requirement of four specific magnitude measurements per redshift
regime significantly reduces the numbers (Table~\ref{venn}).  \setlength{\tabcolsep}{4pt} 
\begin{table*}
  \caption{The number of measurements in each of the main (cf. Fig.~\ref{side}) magnitude bands for the 
optical and radio datasets. The final column gives the  number with 
    $W3\cap W2 \cap\, I\cap U$ divided by the total number expressed as a percentage.}
\centering
\begin{tabular}{l c cccc cc} 
\hline\hline
\smallskip  
  Sample  & No. of & $W3$ & $W2$ & $I$ & $U$ & $W3\cap W2$ & Rate\\
                 &  sources   &          &            &       &           &     $\cap\, I\cap U$ &  [\%]  \\ 
\hline
MgII  & 23\, 659&2793 & 3395 & 6060 &6343 & 1975 &  8.3\\
SDSS          &  50\,000     &17\,230  & 17\,717  &  49\,205  &  48\,522  &   17\,007 &  34.0      \\
LARGESS    & 10\,883 & 326 & 377 & 10\,799 & 9703 & 287 & 2.6\\
VLBI             & 1468     & 140  & 164  & 588 & 608  & 124 & 8.4\\
21-cm         & 819        & 108  & 128  & 452 & 450 & 53  & 6.5 \\
\hline
Total        &   86\,829&  20\,597  & 21\,781  & 67\,104   & 65\,626 &  19\,446 & 22.4\\
\hline
\end{tabular} 
\label{venn}  
\end{table*}
From the table, we see the ``bottleneck'' in $W3\cap W2 \cap\, I\cap U$ for the radio selected sources is due to the WISE
magnitudes. From our testing, however, the $\lambda=4.6$~$\mu$m and 12~$\mu$m far-infrared magnitudes are necessary for
a reliable photometric redshift prediction (Sect.~\ref{sec:samp}).  Substituting the two WISE magnitudes for the Spitzer
5.8~$\mu$m (``$W1.5$'') and 8.0~$\mu$m (``$W2.5$'') values\footnote{The Spitzer 4.5~$\mu$m band is sufficiently close to
  $W2$ to be counted in the above analysis (Sect.~\ref{pm}), which also applies to the 3.6~$\mu$m band
  ($\lambda=3.4$~$\mu$m for $W1$).}, gives $r=0.4718$ ($n = 392$) for $(I-W1.5)/(W2.5-U)$, cf. Fig~\ref{side}
(middle). Substituting $W1.5$ for $W1$, thus restoring some of the $W3$ to $W2$ wavelength span improves upon this,
with $r=0.5594$ ($n = 392$), although the fit is still inferior to that of $(I-W2)/(W3-U)$.

\subsection{Physical interpretation}
\label{pi}

\begin{figure}
\centering \includegraphics[angle=-90,scale=0.52]{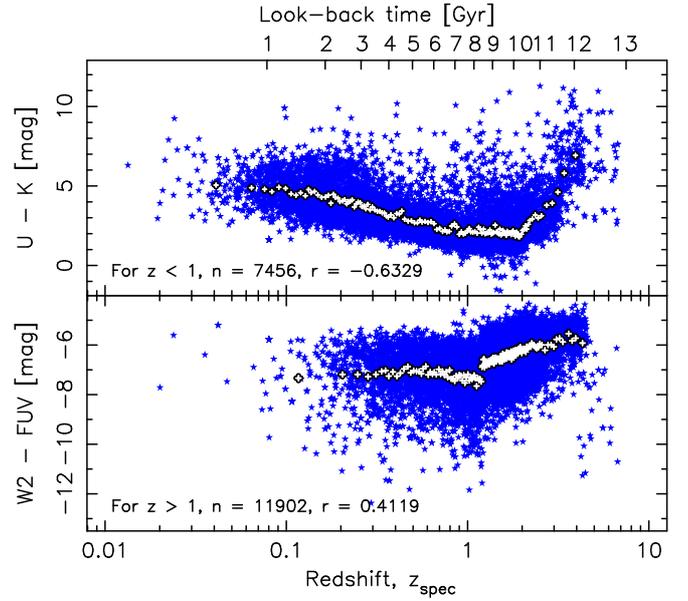}
\caption{The observed $U-K$ and $W2 - FUV$ colours versus redshift for the SDSS sample. The overlain error bars show the mean values in equally sized bins (of 100), spanning the
error range ($\pm1\sigma/10$). In the bottom panel $n = 4322$ and $r = -0.0366$ for $z_{\rm spec} < 1$.}
\label{both}
\end{figure}

It is remarkable that the $(U-K)/(W2 - FUV)$ ratio provides a reliable tracer over such a wide redshift range, provided it
is shifted accordingly to the corresponding observed-frame magnitudes (Fig.~\ref{side}). The near-infrared emission in
QSOs is believed to arise from hot dust in the circumnuclear material heated by the AGN \citep{hos+10}, although
\citet{dvs+02} suggest little contribution to the $K$-band NIR flux. This would suggest a stellar heated dust
component and since the ultraviolet emission is responsible for the excess blue colour of the QSO \citep{shi78,ms82},
an increasing AGN contribution may be apparent as a decrease in $U-K$.  From Fig.\ref{both}, we see that the 
AGN contribution, as traced by $U-K$ does indeed increase with redshift, as expected from the Malmquist bias.
However, for the $W2 - FUV$ colour, which traces a wider range, inverted analogue of $U-K$, we see no correlation
with redshift at $z_{\rm spec} \lapp1$. At higher redshifts, however, where $W2$ approaches $K$ in the rest-frame,
a positive correlation is seen, which may be expected from the $U-K$ anti-correlation. 

This suggests that the $U-K$ colour may be sufficient on its own to estimate the photometric redshift. This was
attempted in order to increase the number of photometric redshifts for the radio sources (Sect.~\ref{maglim})\footnote{Ideally, the photometric redshifts
for radio selected samples would be obtained from the radio photometric properties, although this is proving to be elusive \citep{maj15}.} and returned reasonable values 
($\leq\pm1\sigma$ of $z_{\rm spec}$) over $0.02\lapp z_{\rm spec} \lapp1$. However, as seen in Fig.\ref{both}, there is
a degeneracy between the observed $U-K$ and $z_{\rm spec}$, which requires $W2$ in order to be broken. Hence, we suggest that the
rest-frame $U-K$ colour traces the excess flux due to the AGN and thus offers a measure of the redshift. While
normalisation by $W2 - FUV$ tightens the fit (increasing $|r|=0.63$ to 0.73), the main contribution of this colour
is to flag when the magnitudes should be switched in order to continue to trace the rest-frame $U-K$ emission over a
range of redshifts.

\section{Conclusions}

Given that an uncomplicated, source independent, method of obtaining reliable photometric redshifts will prove
invaluable to the next generation of large  extragalactic radio surveys, we have tested the feasibility of predicting these
using near-infrared and visible magnitudes. This builds upon
the work of \citet{dvs+02}, who find a tight correlation between the $K$-band magnitude and the redshift, although this
only applies to galaxies.  When applied to our initial test sample, which is dominated by bright point sources illuminating
MgII absorption systems (mostly QSOs), we find that the fit of \citeauthor{dvs+02} provides only a
lower limit to the redshift, probably due to an additional contribution (from the AGN) to the near-infrared flux. 

Recently, \citet{gas+18} have estimated photometric redshifts from the WISE $W1$ and $W2$ bands in the 
LARGESS sample. However, with regression coefficients of $r= 0.56$ and 0.36, respectively, the spread
is too wide to yield useful photometric redshift estimates, with poor predictions when applied to other 
datasets. We therefore test how various combinations of magnitudes are correlated with redshift in the 
17\,285 strong test (MgII) sample and find that the ratio of the $I-W2$ and $W3-U$ colours, gives a 
regression coefficient of $r=0.83$ for the 1975 sources for which all four magnitudes were available.
However, expansion of this to a 50\,000 strong sample of QSOs in the SDSS DR12, shows that this
fit fails at redshifts of $z\lapp1$, where the MgII data are scarce. Further testing
finds that the low redshift regime is best fit by the ratio of $U-K$ and $W2-FUV$ colours, which
are essentially the $I-W2$ and $W3-U$ colours in the rest-frame. Likewise, at $z\gapp3$ the correlation
holds best for $(I-W2.5)/(W4-B)$, where $W2.5$ is the Spitzer  8.0~$\mu$m magnitude.

That is, the photometric redshift can be obtained from the rest-frame  $(U-K)/(W2-FUV)$ colour ratio,
over the $0.1\lapp z\lapp5$ span of the data. However, given that we have no a priori knowledge of
the redshift, we estimate this from the (weaker) $W2-z$ correlation, where we find $W2\gapp12.5$ at
$z\gapp1$ and $W2\gapp15$ at
$z\gapp3$. In terms of the observed-frame colours, the photometric redshift is thus obtained from
\[
\log_{10}z_{\rm phot}= \left\{   
\begin{array}{l l}
\frac{1}{0.297}\left(\frac{U-K}{W2-FUV}+0.285\right) & \text{ if }  W2 \leq 12.5  \\ 
\frac{1}{0.376}\left (\frac{I-W2}{W3-U}+0.643\right)  & \text{ if } 12.5 < W2 \leq 15 \\  
\frac{1}{0.849}\left (\frac{I-W2.5}{W4-B}+0.951\right) & \text{ if }  W2 > 15.\\
\end{array}
\right.
\]
Self-testing this on the SDSS sample, the distribution is close to Gaussian with a 
mean $\Delta  z = -0.07$ and a standard deviation of $\sigma_{\Delta z}=0.34$.
On first inspection this does not compete favourably with other studies, which find $\sigma_{\Delta z}\approx0.1$,
although these have $\Delta  z$ wings significantly extended past the Gaussian and are only effective over
limited redshift ranges, even after potential outliers have been removed.
Furthermore, derivation of the photometric redshifts involve complex algorithms in order to break 
the degeneracies which arise via these methods.

Testing our model on the radio sources for which we have redshifts, the $\Delta  z$ distribution is very 
similar to that of the SDSS sample, which gives us confidence in the potential of this method to estimate 
the photometric redshifts of the vast majority of radio sources which lack the required spectroscopic information.
The major drawback is that, although the requirement of four separate magnitude measurements is possible
for $34$\% of the SDSS sources, this falls to 3--8\% in the radio selected samples.
This bottleneck is due to the limited number of WISE magnitudes, with the other magnitudes being available for over half of the  sources (e.g. the $I$ and $U$ magnitudes for $12.5 < W2 \leq 15$, where the vast majority of sources are located). 

Inspection of the colours shows that it is the rest-frame $U-K$ colour which is anti-correlated with redshift, which would
be expected if the ultraviolet emission traces the AGN activity with the far-infrared being dominated by
stellar activity, thus providing an analogue of the $K-z$ relation for galaxies. Reliance upon these two magnitudes
only would vastly increase the applicability of this method as a photometric redshift predictor. However, 
the WISE bands are required to measure $K$ at $z\gapp1$ and $W2$ is required to estimate the redshift 
in order to apply the correct colour combination. Nevertheless, even a 2\% rate will yield photometric redshifts
for over one million of the sources expected to be detected with the Evolutionary Map of the Universe.

\section*{Acknowledgements}

We wish to thank the referee for their helpful comments.
This research has made use of 
the NASA/IPAC Extragalactic Database (NED) which is operated by the Jet Propulsion
Laboratory, California Institute of Technology, under contract with the National Aeronautics and Space
Administration. This research has also made use of NASA's Astrophysics Data System Bibliographic Services.


\begin{appendix} 

\section{Comparison with machine learning methods: \boldmath{$k$}-nearest neighbour}
\label{appa}

\begin{figure}
\centering \includegraphics[angle=-90,scale=0.52]{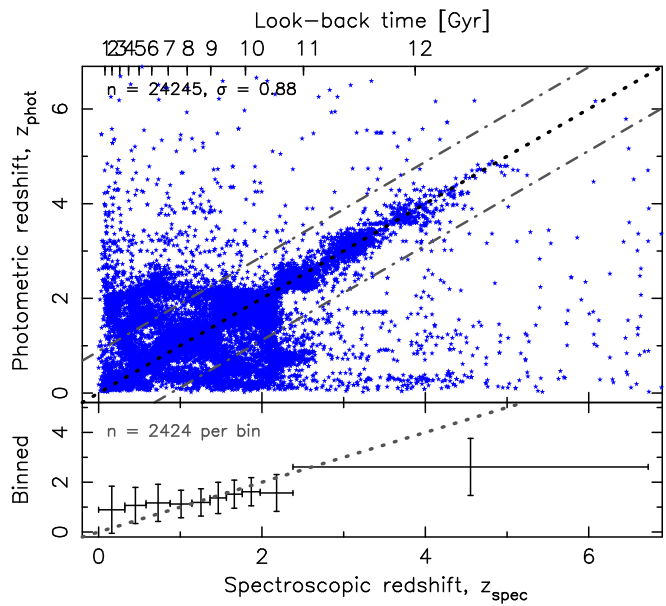}
\caption{The photometric redshifts of the SDSS sample obtained from the $k$-nearest neighbour algorithm, with no source
  filtering.  The standard deviation of $\sigma = 0.8840$ gives a standard error of $\sigma_{\bar{\Delta z}} = 0.0057$
  for $n=24\,245$, cf. $\sigma = 0.5560$ and $\sigma_{\bar{\Delta z}}= 0.0047$ for $n=13\,964$ (Fig.~\ref{lin}).}
\label{kNN_est}
\end{figure}

As stated in Sect.~\ref{cwos}, machine learning techniques are often used to obtain the photometric redshift. One
such method is the $k$-nearest neighbour (kNN) algorithm, which compares the Euclidean distance between a test sample
point and its $k$ nearest neighbours in a feature space, comprising such properties as magnitude, 
colour or luminosity. It then assigns a weighted combination
of the redshifts of those nearest neighbours to the test object in order to place it  into a group. 
This method has been tested extensively on SDSS data, with  the $u- g$, $g - r$, $r - i$ and $ i -  z$ colours 
giving the best results (e.g. \citealt{bbm+08,pzg13,hdzz16}). For our sample of 50\,000 QSOs from the SDSS DR12, this
complete set of colours could be found for 48\,490. Using half the sample to train the model, we obtain
the photometric redshifts shown in Fig.~\ref{kNN_est}.
The distribution has a similar shape to that of \citet{hdzz16}, who 
apply support vector machine methods on top of the kNN algorithm. From the binned data, we see that the
large apparent spread at low redshift is countered by a large population of sources for which 
$z_{\rm phot}\approx z_{\rm spec}$, giving reasonable statistical accuracy at $z_{\rm spec}\lapp2$,
although the mean photometric redshift is underestimated at higher redshifts. 

\begin{figure}
\centering \includegraphics[angle=-90,scale=0.52]{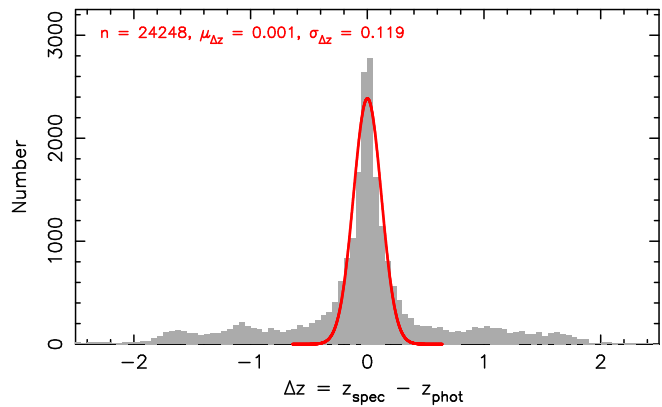}
\caption{The $z_{\rm spec} - z_{\rm phot}$ distribution of the SDSS sample from the kNN algorithm (Fig.~\ref{kNN_est}).}
\label{est_histo}
\end{figure}

In Fig~\ref{est_histo}, we show the corresponding distribution of $\Delta  z = z_{\rm spec} - z_{\rm phot}$.
Although the Gaussian fit is narrow, like the other studies utilising the SDSS colours 
(e.g. \citealt{rws+01,wrs+04,bbm+08,mhp+12}), we find the same wide tails,
resulting in a non-Gaussian (fat/heavy-tailed) distribution. Specifically, the Gaussian fit gives $\mu_{\Delta z} = 0.001$ and  
$\sigma_{\Delta z} = 0.119$, although the data themselves have $\mu_{\Delta z} = -0.011$ and  $\sigma_{\Delta z} = 0.885$.
This compares to a fit of $\mu_{\Delta z} = -0.070$ and   $\sigma_{\Delta z} = 0.345$ using our method, with the 
data giving $\mu_{\Delta z} = -0.174$ and  $\sigma_{\Delta z} = 0.557$ (Fig.~\ref{histo}).

\begin{figure}
\centering \includegraphics[angle=-90,scale=0.52]{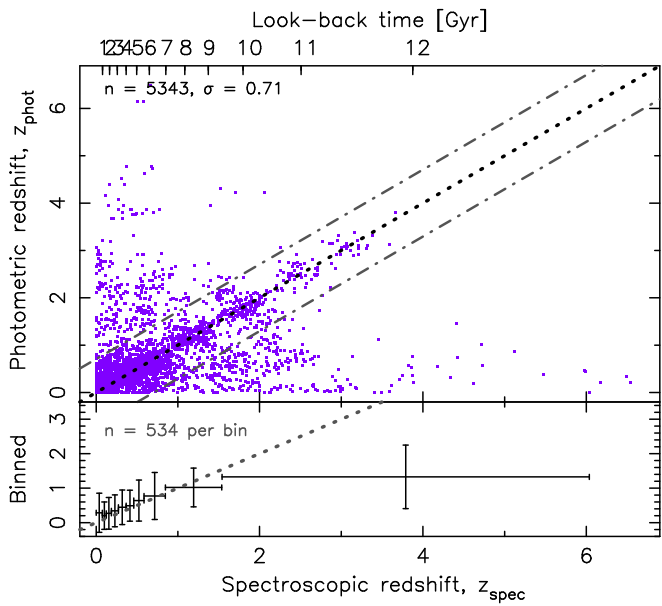}
\caption{The photometric redshifts of the radio source sample obtained from the $k$-nearest neighbour algorithm, with no source filtering.}
\label{kNN_radio}
\end{figure}

\begin{figure}
\centering \includegraphics[angle=-90,scale=0.52]{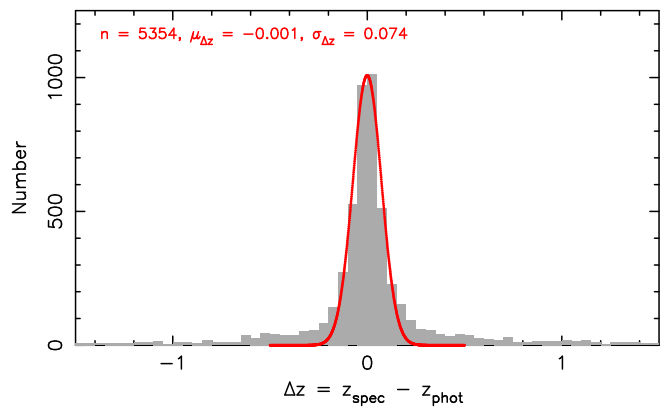}
\caption{The $z_{\rm spec} - z_{\rm phot}$ distribution of the radio source sample from the kNN algorithm (Fig.~\ref{kNN_radio}).}
\label{kNN_radio_histo}
\end{figure}

We also apply the kNN algorithm to the radio selected sources (Sect.~\ref{rss}), where the requirement
of only five (optical) magnitudes, vastly increases the sample (10\,707 cf. our 410). Again, using half
of the data to train the algorithm, we obtain a very similar distribution as the SDSS sample (Fig.~\ref{kNN_radio}),
where, again, this underestimates the redshift at $z_{\rm spec}\gapp2$. Showing the $\Delta z$ distribution (Fig.~\ref{kNN_radio_histo}), 
broad wings are again apparent, with the Gaussian fit underestimating the spread of the data, which have
$\mu_{\Delta z} = 0.009$ and  $\sigma_{\Delta z} = 0.705$. This compares to $\mu_{\Delta z} = -0.255$ and  $\sigma_{\Delta z} = 0.787$ using our method (Fig.~\ref{radio}), which, again,  appears to be more accurate at high redshift (Figs.~\ref{LARGESS_pred} to \ref{21-cm_pred}).

The  machine learning techniques offer a comparable accuracy to our method, 
with the narrow $\Delta z \approx0$ peak being countered by broad wings, whereas we 
obtain a generally wider spread but a more Gaussian distribution. Although the large number
of magnitudes required for our method vastly reduces the sample size, we do not require a 
training set and the colour ratios we employ have a clearer physical interpretation, giving
insight into the AGN contribution to the colour (Sect.~\ref{pi}).  It is thus apparent that our method
provides a useful independent means with which to determine the photometric redshift, thus providing
another string in the bow in determining this for the large sample of 
objects expected from forthcoming large radio surveys.

\end{appendix}

\end{document}